\documentclass[runningheads]{llncs}\usepackage[]{graphicx}\usepackage[]{xcolor}
\makeatletter
\def\maxwidth{ %
  \ifdim\Gin@nat@width>\linewidth
    \linewidth
  \else
    \Gin@nat@width
  \fi
}
\makeatother

\definecolor{fgcolor}{rgb}{0.345, 0.345, 0.345}

\usepackage{framed}
\makeatletter
 {\par\unskip\endMakeFramed%
 \at@end@of@kframe}
\makeatother

\definecolor{shadecolor}{rgb}{.97, .97, .97}
\definecolor{messagecolor}{rgb}{0, 0, 0}
\definecolor{warningcolor}{rgb}{1, 0, 1}
\definecolor{errorcolor}{rgb}{1, 0, 0}
\newenvironment{knitrout}{}{} % an empty environment to be redefined in TeX

\usepackage{alltt}
\usepackage{subcaption}
\usepackage{graphicx}
\usepackage{acronym}
\usepackage{multirow} 
\usepackage{tabularx}
\usepackage{hyperref}
\usepackage{amsmath}

\newcommand{\ie}            {\textit{i.e.}} % i.e.
\newcommand{\eg}            {\textit{e.g.}} % e.g.
\newcommand{\etc}           {\textit{etc.}} % etc.
\newcommand{\rom}[1]        {\textsc{({\romannumeral #1})}} % roman inline lists
\newcommand{\stage}[1]        {\textsc{({S#1})}} % roman inline lists
\newcommand{\repo}        {\url{github.com/sergiomrebelo/evo-poster}} 

\newacro{cd}[CD]    {Computational Design}
\newacro{gd}[GD]    {Graphic Design}
\newacro{ec}[EC]    {Evolutionary Computation}
\newacro{iec}[IEC]  {Interactive Evolutionary Computation}
\newacro{ga}[GA]     {Genetic Algorithm}

\newacro{gan}[GAN]       {Generative Adversarial Network}

\newacro{iec}[IEC]       {Interactive Evolutionary Computation}
\newacro{css}[CSS]       {Cascading Style Sheets}
\newacro{ai}[AI]         {Artificial Intelligence}
\newacro{cv}[CV]         {Computer Vision}
\newacro{api}[API]       {Application Programming Interface}
\newacro{svg}[SVG]       {Scalable Vector Graphics}
\newacro{png}[PNG]       {Portable Network Graphics}
\newacro{json}[JSON]     {JavaScript Object Notation}
\newacro{y}[Y]           {Luminance}
\newacro{nsga2}[NGGA-II] {Elitist Non-dominated Sorting Genetic Algorithm}
\newacro{mit}[MIT]       {Massachusetts Institute of Technology}
\newacro{ann}[ANN]       {Artificial Neural Network}
\newacro{ml}[ML]         {Machine Learning}
\newacro{cc}[CC]         {Computational Creativity}
\newacro{ddd}[DDD]       {Domain-Driven Design}
\newacro{otvf}[OTVF]     {OpenType Variable Font}
\IfFileExists{upquote.sty}{\usepackage{upquote}}{}
\begin{document}

\title{Evaluation Metrics for Automated Typographic Poster Generation\thanks{This paper is based upon work from a scholarship supported by SPECIES, by the Foundation for Science and Technology, I.P./MCTES through national funds (PIDDAC), within the scope of CISUC R\&D Unit - UIDB/00326/2020 or project code UIDP/00326/2020, and by Ministerio espa\~{n}ol de Econom\'{\i}a y Competitividad (Spanish Ministry of Competitivity and Economy) under project PID2020-115570GB-C22 (DemocratAI::UGR).}}

\author{S\'{e}rgio M. Rebelo\inst{1}\orcidID{0000-0002-7276-8727} \and J. J. Merelo\inst{2}\orcidID{0000-0002-1385-9741} \and Jo\~{a}o Bicker\inst{1}\orcidID{0000-0002-0670-1217}\and Penousal Machado\inst{1}\orcidID{0000-0002-6308-6484}}

\institute{University of Coimbra, CISUC/LASI -- Centre for Informatics and Systems of the University of Coimbra, Department of Informatics Engineering, Portugal\\ \email{\{srebelo,bicker,machado\}@dei.uc.pt}\\ \and Department of Computer Engineering, Automatics, and Robotics and CITIC, University of Granada, Spain\\ \email{jmerelo@ugr.es}}

\titlerunning{Evaluation Metrics for Automated Typographic Poster Generation}
\authorrunning{Rebelo \textit{et al.}}

\maketitle           

\begin{abstract}
\acl{cd} approaches facilitate the generation of typographic design, but evaluating these designs remains a challenging task. In this paper, we propose a set of heuristic metrics for typographic design evaluation, focusing on their legibility, which assesses the text visibility, aesthetics, which evaluates the visual quality of the design, and semantic features, which estimate how effectively the design conveys the content semantics. We experiment with a constrained evolutionary approach for generating typographic posters, incorporating the proposed evaluation metrics with varied setups, and treating the legibility metrics as constraints. We also integrate emotion recognition to identify text semantics automatically and analyse the performance of the approach and the visual characteristics outputs.

\end{abstract}

\section{Introduction}
\acl{cd} has revealed a significant potential to transform practices in the field of \acf{gd}, including Typography and Layout domains
\cite{levin2021a,richardson2016a}. This potential involves automating typesetting tasks (\eg~applying text styles, creating tables,~\etc) and facilitating design exploration. Nonetheless, assessing the outcomes of these computational processes remains a challenging task, given that their evaluation relies on subjective design factors. 

In this paper, we present a set of ten heuristic metrics for evaluating typographic designs. These metrics consist of the application of design rules that enable automated assessment of various characteristics of typographic designs, especially posters. Our motivation behind developing these metrics is to streamline and facilitate computational typesetting processes, ultimately leading to faster and more efficient \acs{gd} practices. The proposed set of metrics encompasses the evaluation of designs in terms of their legibility, aesthetic features, and coverage of content semantics. \textsf{Legibility} metrics determine whether all text content is adequately displayed and readable within the design, including the evaluation of \textit{text legibility} and \textit{grid appropriateness}. \textsf{Aesthetic} metrics assess the visual quality of the designs, examining aspects like \textit{alignment}, \textit{balance}, \textit{justification}, \textit{regularity}, \textit{typeface pairing}, and \textit{negative space fraction}. \textsf{Semantic} metrics focus on assessing how the composed text effectively conveys the semantic meaning of the content in terms of \textit{layout} and \textit{typography}.

We study the practical application of these metrics in guiding the creation of typographic poster designs by developing an \acf{ec} approach. In this context, we considered that the primary goal of posters is to fully display their content. This way, we explored a constrained evolutionary methodology where legibility metrics are treated as constraints that the generated outputs must satisfy, while other metrics define the objective value of designs. This approach is inspired by the workflow of traditional typography design processes, as conducted in nineteenth-century print houses. Back then, typographers employed an algorithmic method for typesetting content to fill all the space in a matrix. They use condensed typefaces for lengthy sentences and extended typefaces for shorter ones, while also emphasising the most significant parts of the content typographically \cite{meggs2016a}.

Furthermore, we designed a user interface to support the developed approach, enabling users to input text content and specify the desired visual features of the outputs. To facilitate the generative process, we develop procedures to automatise the text division and to recognise the more semantically significant parts of the content using emotional recognition. The development of this approach follows an agile science methodology, structured around potential user cases and scenarios for the application of the proposed metrics in poster design \cite{merelo2022a}. The code repository for this project is accessible at~\repo. Supplementary materials are  available at \url{cdv.dei.uc.pt/projects/evoposter} (websites visited: 8 November 2023).

We conducted experiments to examine the influence of the proposed evaluation metrics on the evolutionary generation of typographic poster designs. These experiments involved the legibility-constrained evolution of typographic posters using input texts of varying lengths and emotional content. The experiments consisted of three stages, each focusing on the evolution of posters based on either semantic metrics, aesthetic metrics, or a combination of both. The results demonstrate that the proposed metrics effectively guide an evolutionary process, producing finished and legible designs from a variety of text inputs while considering both aesthetics and semantics.

The primary contribution of this paper is the set of metrics for evaluating typographic designs. Other prominent contributions include \rom{1} a functional constrained evolution approach for creating typographic posters; \rom{2} a multi-purpose, domain-driven, and easily understandable representation of poster designs; \rom{3} an investigation into weighted objective function strategies for evaluated designs based on aesthetics and semantics; \rom{4} an exploration of integrating computational design into a poster design field.

The remainder of the paper is organised as follows. The \hyperref[sec:related-work]{next section} provides a review of the related work, specifically focusing on design metrics. The \hyperref[sec:metrics]{third section} comprehensively describes the metrics. The 
\hyperref[sec:experiments]{fourth section} explains our experimental approach, the setup, and the results. The \hyperref[sec:conclusions]{last section} summarises our contributions and outlines potential directions for future research.

\section{Related Work} 
\label{sec:related-work}
Heuristic metrics and measures for visual assessment are frequently used in the computational generation of visuals, especially in the context of \acs{ec}. These metrics have been explored to overcome the limitations of cooperative evolutionary approaches, like \acl{iec} strategies, allowing the incorporation of subjective human-related data into the evolutionary design processes. However, they can lead to user fatigue and inconsistent evaluations \cite{machado2008a}.

Typography and Layout metrics are employed for either generative or optimisation objectives. Geigel and Loui \cite{geigel2000a}~employed a set of design measures, including page balance, spacing, and emphasis, to automatically evolve page layouts for photography albums. Harrington et al.\cite{harrington2004a}~proposed a non-linear layout evaluation measure, combining a set of heuristic metrics for document design, such as alignment, regularity, separation, balance, white-space fraction, white-space flow, proportion, uniformity, and page security.  Building upon these measures, Purvis et al \cite{purvis2003a} and Rebelo et al. \cite{rebelo2021a}~developed generative evolutionary approaches, adapting the proposed measures for the characteristic of the outputs. Lok et al.~\cite{lok2004a}~introduced a technique for computing visual balance using lightness weight maps from images. Balinsky et al.~\cite{balinsky2009a}~presented measures for determining page alignment and regularity, based on the extraction and quantification of ``alignment lines.'' Bylinskii et al.~\cite{bylinskii2017a} and Xie et al~\cite{xie2021a} developed \acf{ml} models which unveiled some potential for auto-completion and layout retrieval. More recently, Lopes et al.~\cite{lopes2023a} developed a pixel-based approach to assess the balance~of~design.

The reviewed related work primarily focuses on visual attributes like balance, regularity, and alignment while overlooking typographic attributes. Nevertheless, these typographic attributes are an essential consideration for graphic designers during their creative processes. Factors like typography pairing and content emphasis significantly contribute to the distinctiveness and effectiveness of the conveyed message. Recent \acs{ml}~approaches show a promising manner to evaluate the designs. However, their effectiveness in poster design scenarios is constrained by the limited availability of poster layout data.

\section{Metrics} 
\label{sec:metrics}
We introduced a set of ten metrics to evaluate typographic designs, specifically posters that feature short text messages. These metrics aim to address the lack of consideration of typography and representation of content semantics in current work on computational assessment of visual designs. 

The computation of these metrics assumes that designs consist of text boxes, each representing a single line of text. These text boxes are characterised by attributes such as size, alignment, and font. They are organised sequentially within a one-column grid. In addition, typefaces used must provide data related to the category,~\eg~serif, mono-space, sans-serif,~\etc~Certain metrics require quantifying the emotional charge of each text line. An optimal poster layout is also considered, which describes the expected distribution of each text box in terms of percentages of the poster's height based on the emotional analysis. 

The metrics are evaluated on a scale ranging from 0 (poor score) to 1 (perfect score), and they can be divided into three evaluation objectives: \textsf{legibility}; \textsf{aesthetics}; and \textsf{semantics}. They have been implemented as a standalone module, and the source code is available in the project repository. 

\textsf{Legibility} metrics evaluate whether the text content on a poster is fully visible in the design. This objective includes two distinct metrics: \rom{1} the \textit{text legibility} and \rom{2} the \textit{grid appropriateness}. 

The \textit{text legibility} metric assesses whether all text content is fully visible within the text boxes that compose the poster. It is calculated by the arithmetic mean of the legibility score of all text boxes. The legibility score of each text box is the variance between the width of the text box's content, as rendered, and the available width of the design (\ie~the width of the poster without the horizontal margins). This score is then normalised, gradually prejudicing the cases when the text exceeds the available width. A score of 0 indicates that the text width doubles the width of the container, while a value of 1 indicates that the text fits within the space. 

The \textit{grid appropriateness} metric assesses whether the grid used in the design is suitable. It does so by comparing the size of the grid (both width and height, including the margins) with the size of the poster. This metric can only have two values. A value of 1 indicates that the grid entirely fits within the design dimensions, while a value of 0 suggests that it does not.

\textsf{Aesthetic} metrics are used to evaluate the visual and typographic features of a design. These metrics include \rom{3} \textit{alignment}, \rom{4} \textit{balance}, \rom{5} \textit{justification}, \rom{6} \textit{regularity}, \rom{7} \textit{typeface pairing}, and \rom{8} \textit{negative space fraction}. We defined these metrics based on works of Harrington et al.~\cite{harrington2004a} and typographic principles outlined by Lupton~\cite{lupton2014a} and Bringhurst~\cite{bringhurst1997a}.

The \textit{alignment} metric assesses the consistency of the horizontal alignment of text. The estimation involves two main steps. First, it computes the variance in text width between neighbouring text boxes. Then, it calculates the arithmetic mean of these variances $(d)$, using the non-linear $A / (A + d)$ where $A$ is a constant that controls how fast values fall away from 1 as the distance between entries increases. Second, the metric checks if the line alignment of the text box is uniform. This is determined by the division of 1 (the expected line alignments on the design) by the count of different line alignments identified on the posters. The overall alignment score is a weighted average, with width variance contributing 80\% and text alignment contributing 20\%. This score is subsequently normalised within a range from 0 (high width variance and different line alignments) to 1 (low width variance and consistent use of a single line alignment).

The estimation of \textit{balance} metric initially involves the calculation of the visual weights and the balance centres of each text box which compose it. The visual weight of a text box $(vw)$ is defined by multiplying its area by its optical density. The optical density $(oD)$ is calculated by considering the relative Luminance, which is computed on the average pixel values for the red $(r)$, green $(g)$, and blue $(b)$ colour channels within the text box, using the formula:
\begin{equation}
\label{eq:optimal-density}
oD = \log_{10} \left( \frac{1}{n} \sum_{n}^{t=1}0.2126\times r_{t} + 0.7152 \times g_{t} + 0.0722 \times b_{t} \right)
\end{equation}
The balance centre of a text box $(x_t, y_t)$ is determined based on its line alignment $(x_t)$ and the overall vertical alignment of the poster. For instance, if the text box is left-aligned and vertically aligned to the top, its balance centre $(x_t, y_t)$ is the upper-left corner. When the vertical alignment is set to centre, the vertical balance centre position is visually adjusted moving one-twelfth towards the top. Next, it calculates the centre of the visual weight of the entire design $(w_x, w_y)$,  by considering all text boxes $(t)$ within it as:
\begin{equation}
\label{eq:centre-visual-weight}
w_x = \left (\frac{\sum_{t=1}^{n}x_t \times vw_t}{\sum_{t=1}^{n}vw_t}  \right) 
\: \text{and} \: 
w_y = \left (\frac{\sum_{t=1}^{n}y_t \times vw_t}{\sum_{t=1}^{n}vw_t}  \right)
\end{equation}
Subsequently, it estimates the expected balance centre of the design $(c_x, c_y)$ employing the same method as used for text boxes, considering the alignment of the first text box $(c_x)$ and the same global vertical alignment $(c_y)$, albeit taking into consideration the full poster size. Finally, the overall balance $(B)$ score is determined considering the calculated current and expected balance centres and poster sizes as follows:
\begin{equation}\label{eq:balance-value}
B = 1 - \left [ 
\left (
\left ( \frac{w_x - c_x}{width} \right )^2 + 
\left ( \frac{w_y - c_y}{height} \right )^2
\right ) / 2
\right ]^{\frac{1}{2}}
\end{equation}

The \textit{justification} metric evaluates whether the text fully occupies the available space. This metric is inspired by the traditional aesthetics of nineteenth-century letterpress posters where text content was traditionally justified within the available space, ensuring that text occupies all the available areas when possible. The calculation of this metric is similar to the \textit{text legibility}. The overall justification score is determined as the arithmetic mean of the justification scores of all text boxes. The justification score of a text box is calculated by considering the variance between the text width of each text box and the available space. The variance value is normalised penalising both designs with text overflow and with excessive white space. The penalty is lower when text fully fits within the poster, being divided by a factor. A lower variance results in a higher justification score. A justification score of 1 indicates zero width variance, while if the variance doubles the design width the justification is set to 0.
 
The  \textit{regularity} metric evaluates how regular is text box heights in the design. It measures the distances $(d)$ between the vertical positions of the top edges in neighbouring text boxes using a non-linear function $A / (A + d)$, where $A$ is a constant that determines the rate at which values decrease as the distance between entries increases, similar to \textit{alignment}. The overall regularity score is the arithmetic mean of the value of all neighbouring pairs is normalised within a range between 0 (low regularity) and 1 (high regularity).

The \textit{typeface pairing} metric evaluates the compatibility of the typefaces used on the poster, considering their categories on typographic classification. It begins to create a unique list of categories from the typefaces used in the design. The overall typographic pairing score is determined by the count of categories found in the design, normalised between 1 (if there is only one category) and 0 (if all the used typefaces are from different categories).

The \textit{negative space fraction} metric assesses the appropriateness of the percentage of background colour in the design. It calculates the current percentage of the poster occupied by background colours and computes the deviation from an optimal value chosen by the user. The differences are then normalised to a range between 0 (twice the optimal background percentage) and 1 (optimal background percentage).

\textsf{Semantic} metrics assess whether the placement of the typography in the designs conveys the semantic meaning of the content. This analysis encompasses the evaluation of the semantic significance of the \rom{9} \textit{layout}, and \rom{10} \textit{typography}.

The \textit{semantic significance of the layout} metric evaluates whether text lines which contain more emotional charge are appropriately highlighted in the layout. The greater the emotional charge on a line, the more height this line of text should be typeset on the design. This calculation is done at the line (or text box) level. So, for each text box, it calculates the distance between its height and the height in the optimal layout, represented as a percentage of the poster's total height. This metric can operate in two modes, ``Fixed'' and ``Relative''. In the ``Fixed'' mode, it considers that content must fulfil the total poster available height. In ``Relative'' mode, it only considers the current height of the composition, ignoring the empty space. By default, it operates in ``Fixed'' mode. The overall score for the semantic significance of the layout is the arithmetic mean of the normalised distance of all text boxes, ranging from 0 (significant differences between the actual and optimal layout) to 1 (perfect match between the optimal and current layout).

The \textit{semantic significance of typography} metric evaluates if the most emotional parts of the content are typographically emphasised. This involves calculating variations in weight, stretch, and type design across fonts used in the text boxes within the design. The score for the variable typographic features (weight and stretch) is computed as the mean of the distances between the current and expected values. These distances are normalised between 0 (maximum distance) and 1 (no distance). To determine the expected values for each text box, the range of values used in the design is established. Then, for each text box, an expected distance is assigned based on the recognised emotional levels in the content. The text box with the highest emotional context recognised is assigned the maximum distance value within the range, while the one with lesser emotional charge receives no distance. Ultimately, for each text box, the distance between the current value and the value of a text box with no expected difference is calculated, and the current distance is determined by comparing the results with the assigned expected distance. The type design score is calculated by counting the number of typefaces that deviate from the expected ones, normalised based on the total number of text boxes. A score of 0 indicates that the output aligns with type design expectations, whereas a score of 1 signifies that it does not meet any expectations. To calculate the expected typefaces, typefaces are initially assigned to emotional levels based on their first use within each emotional level. Each typeface is uniquely associated with one emotional level. If a typeface is already assigned to another level, the typeface for this level is set as undefined. The overall score of the \textit{semantic significance of typography} is the highest score among the three types of variations. To prevent excessive emphasis on the same content, the score is penalised when multiple features receive evaluations exceeding a high threshold. In such cases, the score is divided by the number of features that exceeded this threshold.

\section{Experiments} 
\label{sec:experiments}
We conduct evolutionary experiments using the proposed metrics and a \acl{ga} to automatically evolve typographic posters using different texts. We experiment with a constrained evolution approach employing a stochastic ranking method \cite{runarsson2000a} to fitness assignment the designs. \textsf{Legibility} metrics serve as constraints that the generated outputs must adhere to. \textsf{Aesthetic} and \textsf{semantic} metrics determine the design value of the outputs, through a weighted multi-criteria objective evaluation function. In addition, we implement a constraint penalty approach to facilitate the sorting and visualisation of population and elitism. The penalty is calculated as the inverted weighted arithmetic mean of the two \textsf{legibility} metrics.

We adopted a \acl{ddd}~\cite{evans2015a} approach to define the representation of posters. Each genotype is encoded in the \acs{json} format, with each representational attribute identified by a key term commonly used in the \acs{gd} domain. The genotype comprises two parts: \rom{1} text boxes data and characteristics, namely content, typeface, weight, stretch, size, and alignment; and \rom{2} poster characteristics, specifying the size (width and height), margins (in percentages of size), and vertical alignment. 
The grid of the poster is inferred based on the text box size and the poster characteristics. One descriptive example of a genotype is provided in supplementary materials.

The generative process starts with the random initialisation of a poster population. For each line of text, a text box is assigned, and its characteristics are randomly determined from available choices. This process involves randomly selecting a typeface, along with the weight and stretch based on its available options. The font size is defined within a predefined range, while alignment is randomly set to left, centre, or right. The vertical alignment of the poster is also randomly defined as top, middle, or bottom.

Tournament selection is used to choose individuals for breeding based on their fitness, and variation operators (mutation and crossover) are applied to the selected individuals, with a certain probability, to generate new offspring. We use a uniform crossover method \cite{syswerda1989a}, which randomly determines which parent will pass down genetic material to the descendant. It flips the vertical alignment gene and all text boxes on the new individual to determine which parent will pass down its characteristics to the offspring. The text boxes are passed to the descendants with all of their characteristics. 

The mutation variation operator determines if an attribute of the individual in new offspring will undergo change, based on a certain probability. The mutation method employed varies depending on the selected attribute. When poster vertical alignment or the alignment, weight, and stretch attributes of a text box are selected for mutation, the current value is modified by one of the available options. If the typeface attribute of the text box is selected, it replaces the current typeface with one of the available options, and it adjusts the weight and stretch attributes if needed, selecting values close to the current ones. In the case of the font size of the text box, it adjusts the size based on random values within a specific range, typically between -5 and 5. 

We employ an elitist process, where the best individuals from the current generation are combined with the new offspring. To select these individuals, it sorts the population using a penalty constraint approach. 

Before the evolution, the system emotionally analyses input content to determine the parts that must be emphasised on the layout. The analysis starts with the preprocessing of the text by \rom{1} handling contracted word forms, \rom{2} replacing abbreviations and slang expressions with formal equivalents, \rom{3} substituting emojis with their actual meanings, \rom{4}  changing negations to antonyms, \rom{5} eliminating stop words and URLs, and \rom{6} tokenising the text. Subsequently, it conducts a lexicon-based analysis using a word-emotion association lexicon \cite{mohammad2010a}. Based on the result, emotional scores are assigned to individual words. The emotional scores for each line are calculated by summing the scores of words~on~it. 

We designed a user interface that allows users to input the text and configure the desired visual features of the outputs and the evolutionary approach. Furthermore,  new typefaces, options, and default evolution parameters through a dedicated configuration file. Typefaces are incorporated into the system in the \acl{otvf} format. Users can input content either divided into lines or choose to let the system automatically divide the text. Text division is performed using a Sentence Boundary Detection algorithm \cite{reynar1997a} to split the text into sentences and then divide the longer sentences based on a random factor within a predetermined line size range. Screenshots of the developed interface are available in the supplementary material.

% In the next subsections, we outline the experimental setup and present the results obtained. The source code of this approach is accessible in the project repository.

\begin{table}[t]
\centering
\caption{Metrics weights in the respective evaluation objective.}\label{tab:experimental-weights}
\begin{tabular}{|l|l||l|l|}
\hline
\multicolumn{2}{|l||}{Aesthetic metrics} & \multicolumn{2}{|l|}{Semantic metrics} \\
\hline
\hline
Alignment &  10\% & Layout &  50\% \\
Regularity &  10\% & Typography &  50\%\\
Balance &  20\% & &  \\
Negative Space Fraction &  20\% & & \\
Justification &  30\% & & \\
Typeface Pairing &  10\% & & \\
\hline
\end{tabular}
\end{table}

\begin{table}[t]
\centering
\caption{Experimental Parameters.}\label{tab:experimental-setup}
\begin{tabular}{|l|l|l|}
\hline
Type & Parameter & Value \\
\hline
\hline
\multirow{5}{*}{Evolutionary} & Generations &  400 \\
& Population size &  30 \\
& Elite size &  1 \\
& Crossover probability & 90\% \\
& Mutation probability & 10\% \\
& Tournament size & 10 \\
\hline
\multirow{3}{*}{Posters' features} & Poster size (px) & $141\times100$\\
& Default margins (ltrb) & 5\% 5\% 5\% 5\% \\
& Min colour contrast & 2.5 \\
\hline
\multirow{2}{*}{Aesthetic metrics} & Alignment/regularity A constant & 10 \\
& Justification factor & 3 \\
& Optimal negative space factor & 50\% \\
\hline
\multirow{2}{*}{Semantic metrics} & Significance of layout mode & Fixed \\
& Significance of typography threshold & 0.2 \\
\hline
\end{tabular}
\end{table}

\subsection{Experimental Setup}
The conducted experiments are divided into three different stages: \stage{1}~evolving only considering \textsf{semantic} objective; \stage{2}~evolving only considering \textsf{aesthetic} objective; and \stage{3}~evolving considering both objectives, with equal weight on objective evaluation. The score of each evaluation objective is determined as the weighted arithmetic mean of the score of metrics that compose it. The weight of each metric on the objective function is defined empirically, with the goal of creating poster designs inspired by the aesthetics of nineteenth-century letterpress posters, which was one of the main motivations for the development of this work. Table~\ref{tab:experimental-weights} displays the weight assigned to each metric in both the semantic and aesthetic groups.

The conducted experiments occur on the client side of a Chromium web browser. We loaded 8 \acl{otvf}, all of which have weight and stretch axes available. The experimental parameters for these experiments were empirically defined and are summarised in Table~\ref{tab:experimental-setup}. In the three evaluation stages, we generated typographic posters for 20 different texts in three languages (English, Portuguese, and French). These texts expressed various purposes, emotional contexts, and lengths. On average, each input text contained 39.55 characters, divided into 7.15 words, with a maximum of 83 characters and a minimum of 17. The average text line length was 4.15 lines. The selected texts, apart from being neutral, were classified by representing emotions such as anticipation, disgust, joy, sadness, surprise, and trust. Further description of the text and the typographic available options are provided in the supplementary material.

\subsection{Experimental Results}
The experimental results demonstrate that the proposed metrics and approach enable the evolution of legible designs for various types of text inputs. Analysing the progression of constraint penalty in the population (see Fig.~\ref{fig:res-leg}), we observed that the proposed approach leads to a faster reduction in the number of individuals that do not comply with the legibility constraints, either in the best individuals or the population average. Individuals evaluated with maximum fitness are present in the population but do not are legible. A detailed analysis of \textit{grid appropriateness} metric reveals that the proposed approach can consistently generate and maintain individuals that use valid grids. On the other hand, the evolution of \textit{text legibility} is inversely proportional to the constraint penalty value. The results also reveal that the fittest individuals at the end of evolution are always individuals with all text legible.

\begin{knitrout}
\definecolor{shadecolor}{rgb}{0.969, 0.969, 0.969}\color{fgcolor}\begin{figure}[t]

{\centering \includegraphics[width=4.8cm,height=3.2cm]{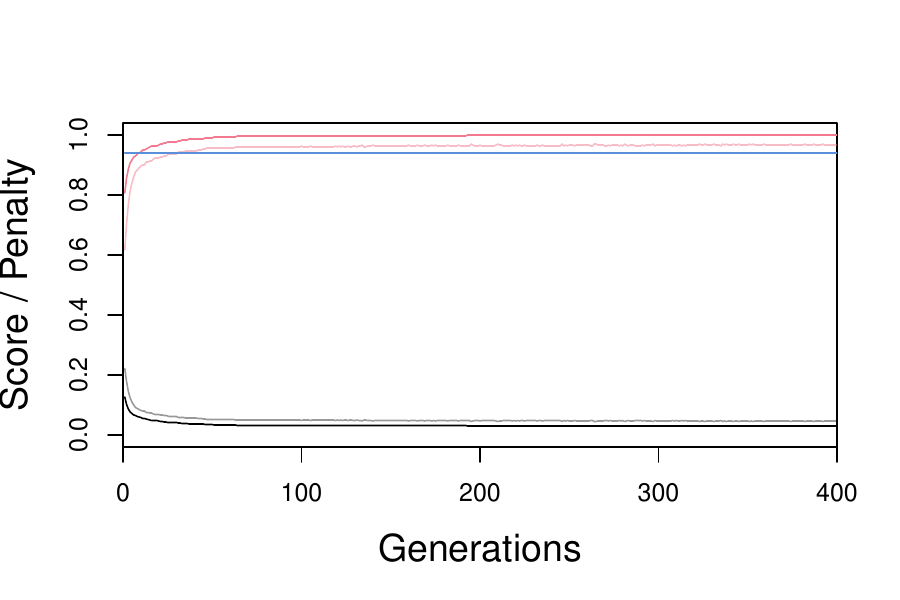} 

}

\caption[Progression of constraint penalty and legibility metrics across generations]{Progression of constraint penalty and legibility metrics across generations. Visualised results present the average of three stages, totalling 1,800 runs. Black lines depict penalty constraint values, while pink and blue lines show the scores of \textit{text legibility} and \textit{grid appropriateness} metrics, respectively. Fittest individuals are depicted with solid lines, and the average population with shaded lines.}\label{fig:res-leg}
\end{figure}

\end{knitrout}

The evolution based on \textsf{semantic} metrics \stage{1} reveals that the proposed semantic metrics can guide the evolution of posters towards designs where the more emotional parts of the content are highlighted. The experimental results (see Fig.~\ref{fig:res-sem}) unveil that the best individual in the population achieves high fitness values as well as high scores in both metrics. However, the metrics values stabilise after a dozen generations, primarily due to the legibility constraints. These constraints restrict the semantic evolution of certain individuals, as they compromise legibility. In this sense, some individuals with high semantic scores are also not fully legible designs. It is also noted that the evolution of the semantic significance of layout is slightly faster than the typography. 

\begin{knitrout}
\definecolor{shadecolor}{rgb}{0.969, 0.969, 0.969}\color{fgcolor}\begin{figure}[t]

{\centering \includegraphics[width=4.8cm,height=3.2cm]{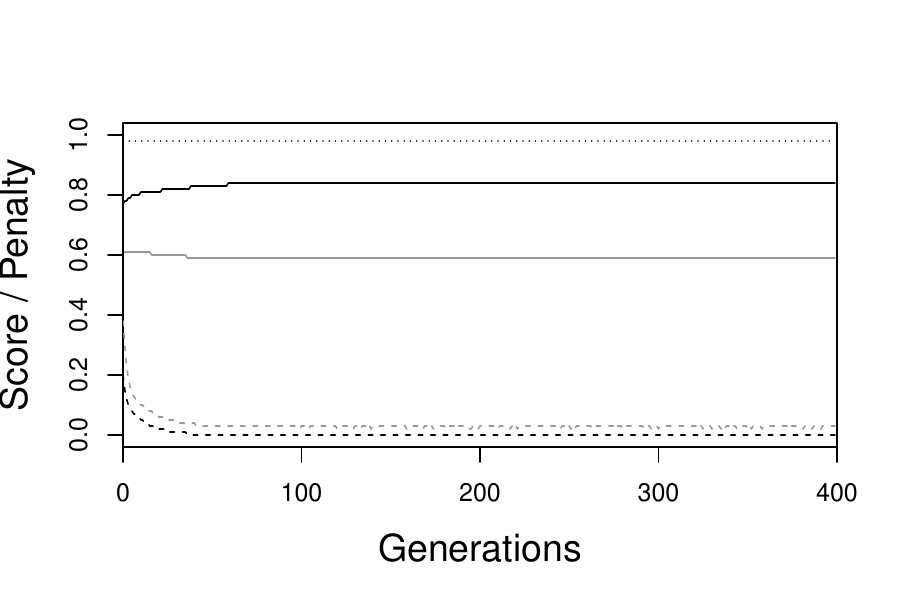} 
\includegraphics[width=4.8cm,height=3.2cm]{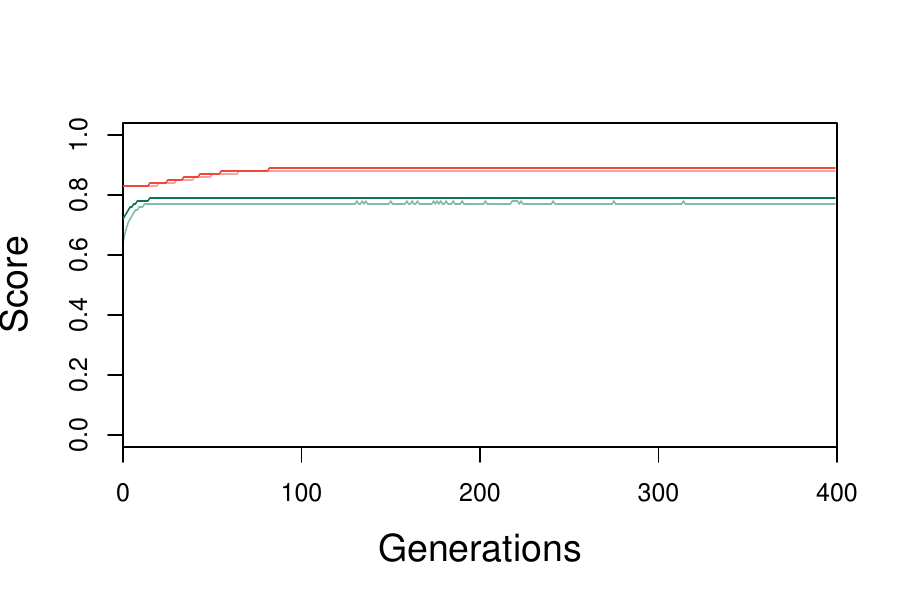} 

}

\caption[Evolutionary progress in the S1 experiment]{Evolutionary progress in the S1 experiment. The results are averages from 600 runs, considering varied text inputs. The chart on the left illustrates best individual fitness (solid lines), constraint penalty (dashed lines) and maximum fitness (dotted). On the right, the chart shows score metrics related to the significance of semantics in \textit{layout} (green lines) and \textit{typography} (orange lines). Solid lines represent the average of the fittest individuals, while shaded lines represent the average of the entire population.}\label{fig:res-sem}
\end{figure}

\end{knitrout}

\begin{figure}[t]
\centering
\label{fig:outputs}
    \begin{subfigure}[h]{1\textwidth}
        \centering
        \includegraphics[width=0.7\columnwidth]
        {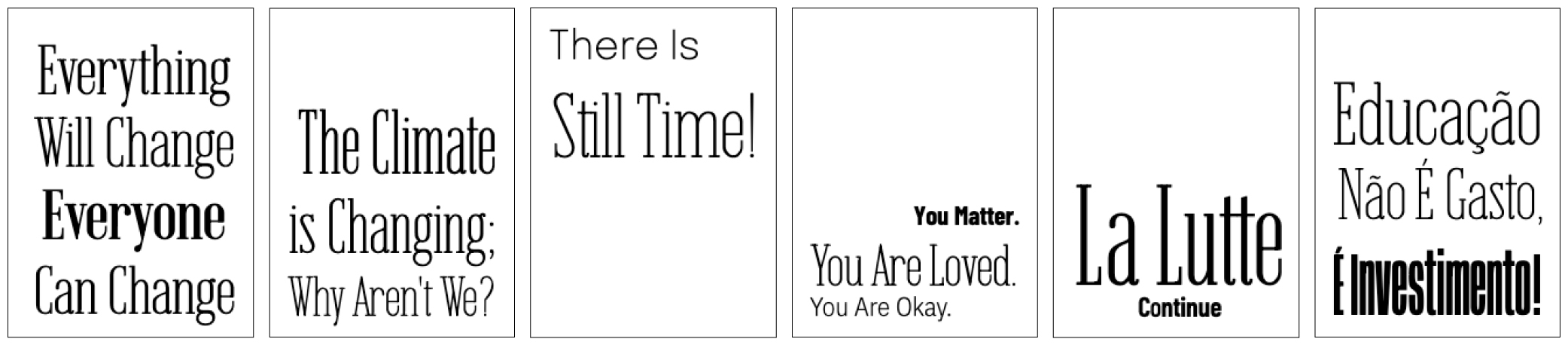}
        \caption{S1 Stage}
        % \vspace{1.5mm}
        \label{fig:s1-outputs}
    \end{subfigure}
    \begin{subfigure}[h]{1\textwidth}
        \centering
        \includegraphics[width=0.7\columnwidth]
        {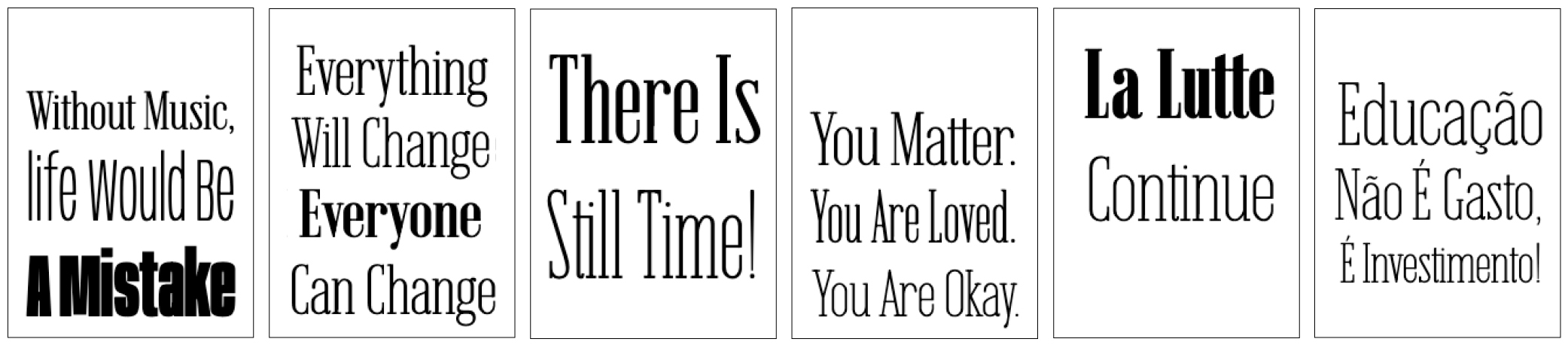}
        \caption{S2 Stage}
        % \vspace{1.5mm}
        \label{fig:s2-outputs}
    \end{subfigure}
    \begin{subfigure}[h]{1\textwidth}
        \centering
        \includegraphics[width=0.7\columnwidth]
        {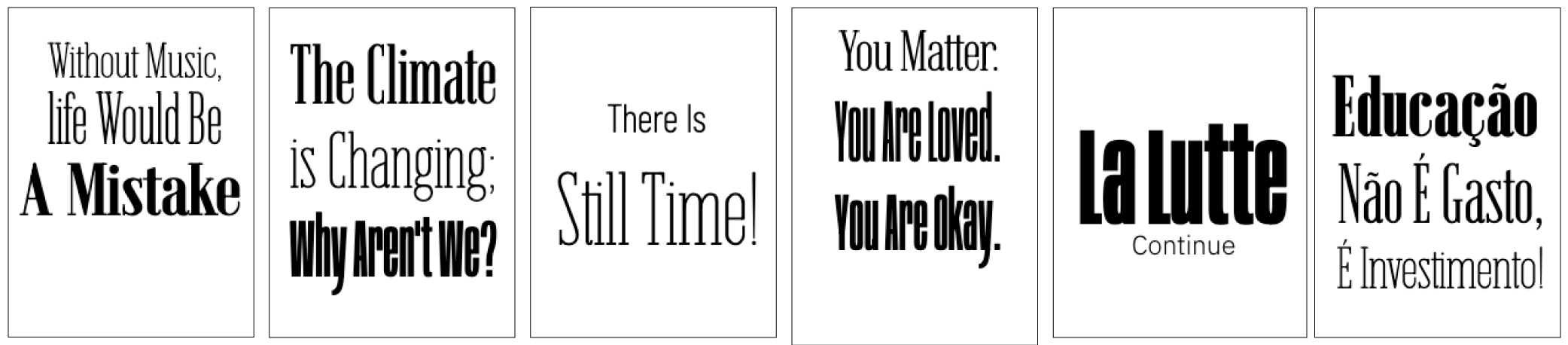}
        \caption{S3 Stage}
        % \vspace{1.5mm}
        \label{fig:s3-outputs}
    \end{subfigure}
    \hspace{5mm}
    \caption{Examples of designs evolved during the three experimental stages. More results can be found in the supplementary material folder.}
\end{figure}

The generated outputs exhibit certain parts of the content that are emphasised in terms of layout, typography, or both (see Fig.~\ref{fig:s1-outputs}). This emphasis is primarily directed toward the more emotional sections of the content. However, when the emphasis affects both layout and typography, it can sometimes become excessive, leading to a visually heavy impact, particularly in designs with fewer emotional variations. Additionally, certain design issues were observed in the outputs, including non-uniform alignment, unbalanced layout, or excessive use of different typefaces.

The outcomes of the \textsf{aesthetics} evolution~\stage{2} indicate that it is possible to evolve designs complying with the proposed aesthetic metrics. Nonetheless, these results also reveal the challenge of evolving designs that fulfil all the metrics simultaneously. The performance of metrics depends on the text content, and the necessity for legibility limits the range of aesthetic design possibilities.

Experimental results (see Fig.~\ref{fig:res-aest}) reveal that global fitness increases faster in the initial generations but slows down as the constraint penalty decreases. We noted that some individuals in the population achieve higher than the best individuals in the population, but they are not legible posters. When examining the score of each metric individually, \textit{regularity}, \textit{alignment}, and \textit{typographic pairing} progressively increase over the generation both in the fittest individuals and population average. This suggests that enhancing these metrics does not compromise the legibility of the posters. On the other hand, \textit{balance}, \textit{justification}, and \textit{negative space fraction} maintain relatively stable scores over the generations, with slight decreases in scores on the best individuals, while legibility and other aesthetic metrics increase. The \textit{justification} score of the best individual was even, on average, below the population average. Further experimentation is necessary to understand how other metrics and experimental parameters impact their evaluation. However, high scores on these metrics result in the creation of invalid individuals, even in more advanced generations, who are subsequently removed from the population. Due to this, legibility metrics had worse scores than in the S1 stage.

\begin{knitrout}
\definecolor{shadecolor}{rgb}{0.969, 0.969, 0.969}\color{fgcolor}\begin{figure}[t]

{\centering \includegraphics[width=4.8cm,height=3.2cm]{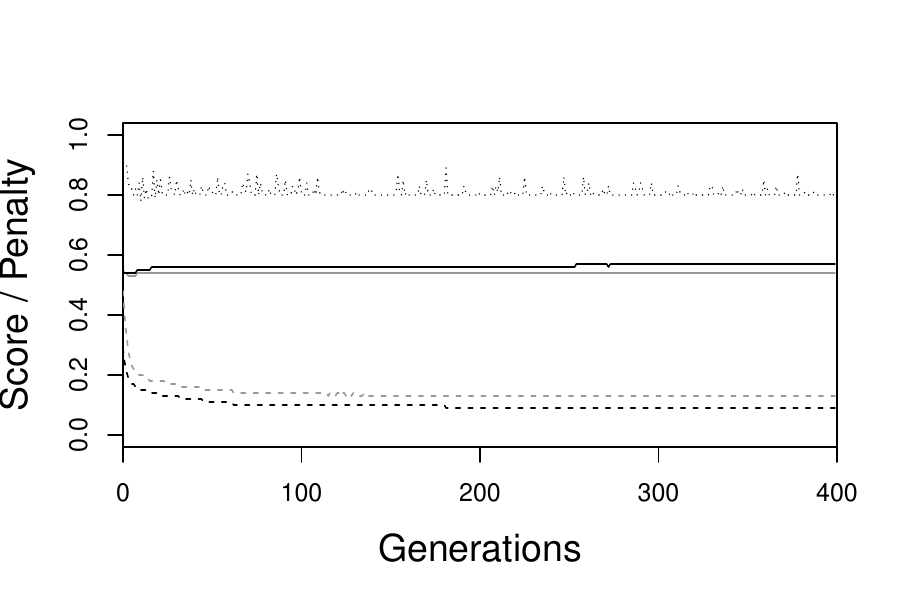} 
\includegraphics[width=4.8cm,height=3.2cm]{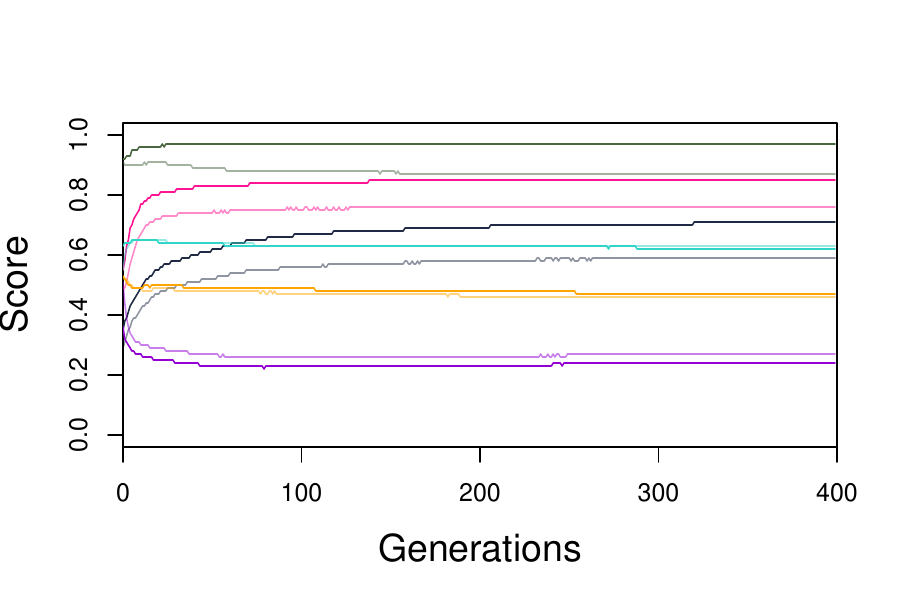} 

}

\caption[Evolutionary progress in the S2 experiment]{Evolutionary progress in the S2 experiment. The results are averages from 600 runs, considering multiple text inputs. The chart on the left illustrates best individual fitness (solid lines), constraint penalty (dashed lines) and maximum fitness (dotted). On the right, the chart shows score metrics related to aesthetics, including \textit{alignment} (navy), \textit{regularity} (green), \textit{justification} (violet), \textit{typography pairing} (pink), \textit{balance} (turquoise), and \textit{negative space fraction} (orange). Solid lines represent the average of the fittest individuals, while shaded lines represent the average of the entire population.}\label{fig:res-aest}
\end{figure}

\end{knitrout}

The visual analysis of the outputs (see Fig.~\ref{fig:s2-outputs}) unveils that, even though they are created with the same input texts, the generated posters are more balanced and regular in typographical terms compared with S1 outputs. Nonetheless, we noticed that in some outputs less meaningful parts of the content are typographically highlighted (\eg~conjunctions), making the content more difficult to read.

The results of the evolution combining both \textsf{aesthetic} and \textsf{semantic} metrics \stage{3} indicate that optimising all the metrics simultaneously is a challenge, being the design possibilities influenced by the text content and the recognised emotional charge. The experimental results (see Fig.~\ref{fig:res-both}) exhibit an evolutionary pattern akin to the two other experiments: faster evolution in early generations followed by gradual stabilisation as the constraint penalty values decrease. However, we noted that there are individuals with higher fitness values in the population, yet they are not legible.

\begin{knitrout}
\definecolor{shadecolor}{rgb}{0.969, 0.969, 0.969}\color{fgcolor}\begin{figure}[t]

{\centering \includegraphics[width=4.8cm,height=3.2cm]{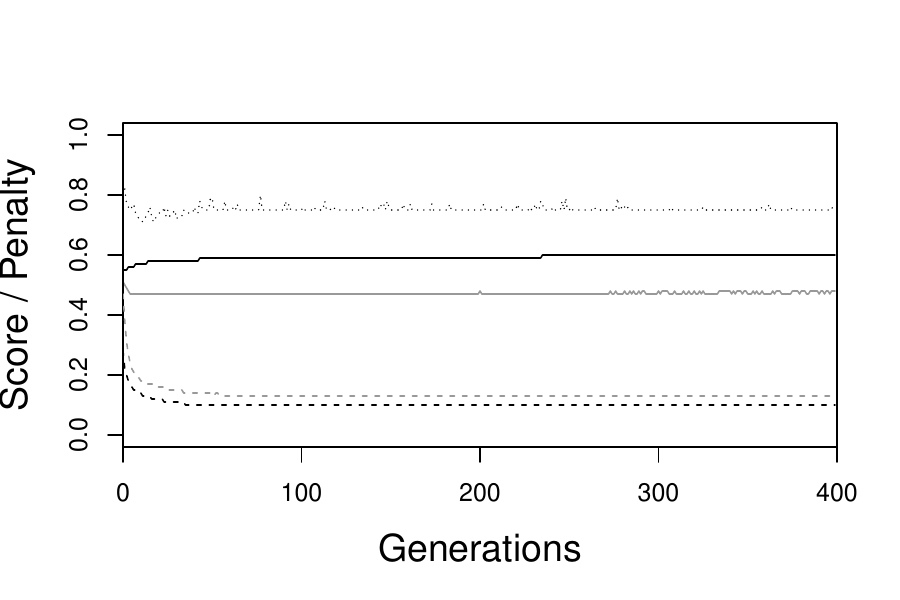} 
\includegraphics[width=4.8cm,height=3.2cm]{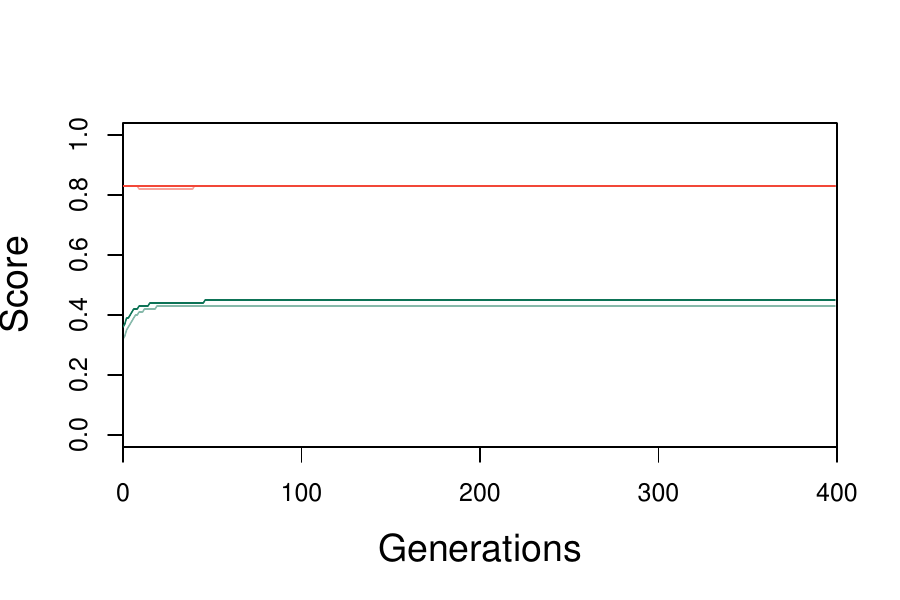} 
\includegraphics[width=4.8cm,height=3.2cm]{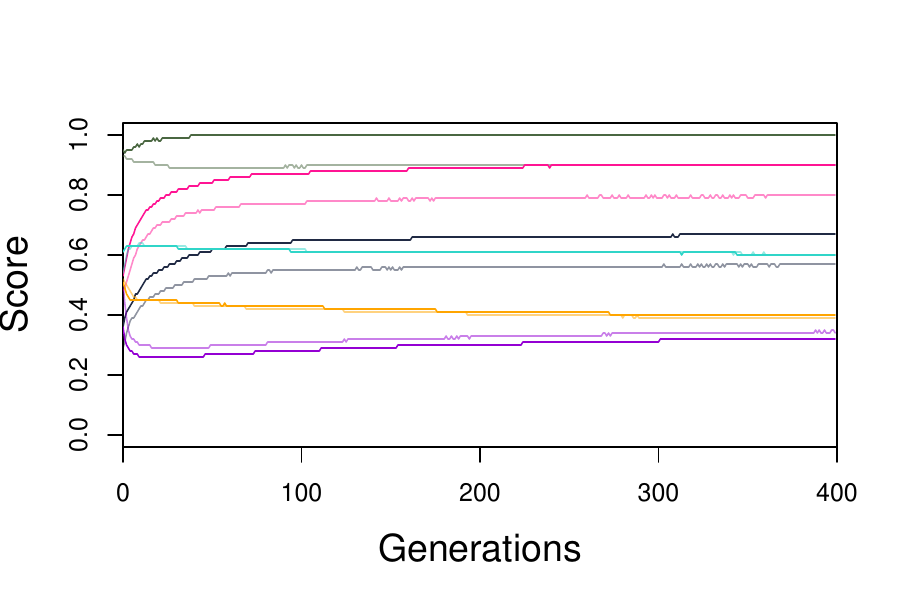} 

}

\caption{Progression of evolution in the S3 experiment. The results are averages from 600 runs, considering various text inputs. The top-left chart presents the progression of the best individual fitness (solid lines), constraint penalty (dashed lines) and maximum fitness (dotted). The top-right and bottom charts illustrated the scores of metrics related to semantics and aesthetics, respectively. Semantic metrics include the significance of semantics in \textit{layout} (green) and \textit{typography} (orange). Aesthetic metrics include \textit{alignment} (navy), \textit{regularity} (green), \textit{justification} (violet), \textit{typography pairing} (pink), \textit{balance} (turquoise), and \textit{negative space fraction} (orange). Solid lines represent the average of the fittest individuals, while shaded lines represent the average of the entire population.}\label{fig:res-both}
\end{figure}

\end{knitrout}

The analysis of the evolution of \textsf{semantic} metrics separately reveals that their combined evaluation with aesthetic metrics presents challenges to evolving semantic metrics, being constrained by the other aesthetic metrics. In terms of the \textit{significance of layout}, this experiment indicates that it is restricted by aesthetic metrics, showing a relatively flat trajectory, with the fittest individuals and the population average achieving similar values. The evolution of the \textit{significance of typography}, although exhibiting some improvement in the early generations, it faster stabilises and remains below the scores achieved in the S1 experiment. The results of \textsf{aesthetic} metrics resemble the results of S2. \textit{Alignment}, \textit{regularity}, and \textit{typographic pairing} progressively increase over the generations, while \textit{balance}, \textit{justification}, and \textit{negative space fraction} maintain relatively stable scores, seemingly constrained in their progress by legibility metrics. We observed a slight evolution of \textit{justification} scores, which appear to be related to the inclusion of semantic metrics and the exploration of layouts that are not highly valued by other aesthetic metrics but, from a semantic perspective, are considered superior and enable improved content justification.

The empirical visual analysis of the outputs (see Fig.~\ref{fig:s3-outputs}) reveals an increase in aesthetic features value present, compared to S1. Simultaneously, it is possible to observe the emphasis of parts with more emotional weight, whether through layout or font selection, in comparison to S2. However, this emphasis in certain designs should be more pronounced in some posters.

\section{Conclusions and Future Work} 
\label{sec:conclusions}
Computational approaches in \acs{gd} unveil potential, but evaluating outcomes remains challenging due to reliance on subjective factors. We introduced ten heuristic metrics for autonomously evaluating typographic designs, addressing both visual and semantic aspects. These metrics are categorised into three evaluation objectives: \textsf{Legibility} metrics assess text display effectiveness, including \textit{text legibility} and \textit{grid appropriateness}. \textsf{Aesthetic} metrics evaluate visual quality, considering \textit{alignment}, \textit{balance}, \textit{justification}, \textit{regularity}, \textit{typeface pairing}, and \textit{negative space fraction}. \textsf{Semantic} metrics focus on typographic content representation for \textit{layout} and \textit{typography}. We applied these metrics in a constrained evolution approach to generate typographic poster designs for various texts, with legibility metrics as constraints and other metrics in a weighted multi-criteria objective evaluation function. Our preliminary experiments covered three stages, evaluating posters using semantic, aesthetic, and both metrics. 

The main motivation behind developing these metrics is to streamline computational typesetting processes. Despite their preliminary nature, evaluation experiments reveal their effectiveness in guiding typographic poster design generation for diverse text contents, highlighting their potential for automation in \acs{gd} practices. However, evolving multiple metrics simultaneously, in this constrained setting, restricts metrics' progression, necessitating further research to understand the individual metrics' progression and their interaction with constraints. Future work also includes less constrained evolutionary experiments with different parameters. Moreover, we intend to conduct further evaluation studies with human participants, aiming to assess the visual impact of the proposed metrics in output designs.

\bibliographystyle{splncs04}
\bibliography{references}

\begin{thebibliography}{10}
\providecommand{\url}[1]{\texttt{#1}}
\providecommand{\urlprefix}{URL }
\providecommand{\doi}[1]{https://doi.org/#1}

\bibitem{balinsky2009a}
Balinsky, H.Y., Wiley, A.J., Roberts, M.C.: Aesthetic measure of alignment and
  regularity. In: Borghoff, U.M., Chidlovskii, B. (eds.) DocEng '09:
  Proceedings of the 9th ACM symposium on Document Engineering. pp. 56--65. ACM
  (9 2009). \doi{10.1145/1600193.1600207}

\bibitem{bringhurst1997a}
Bringhurst, R.: The Elements of Typographic Style. Hartley \& Marks Publishers,
  2nd edn. (1997)

\bibitem{bylinskii2017a}
Bylinskii, Z., Kim, N.W., O'Donovan, P., Alsheikh, S., Madan, S., Pfister, H.,
  Durand, F., Russell, B., Hertzmann, A.: Learning visual importance for
  graphic designs and data visualizations. In: Gajos, K., Mankoff, J.,
  Harrison, C. (eds.) Proceedings of the 30th Annual ACM Symposium on User
  Interface Software and Technology. pp. 57--69. ACM (10 2017).
  \doi{10.1145/3126594.3126653}

\bibitem{evans2015a}
Evans, E.: Domain-Driven Design Reference: Definitions and Pattern Summaries.
  self-publishing (2015), \url{https://www.domainlanguage.com/ddd/reference},
  (visited: 18 January 2024)

\bibitem{geigel2000a}
Geigel, J., Loui, A.C.P.: Automatic page layout using genetic algorithms for
  electronic albuming. In: Beretta, G.B., Schettini, R. (eds.) Internet Imaging
  II. vol.~4311, p. 79 – 90. SPIE (2000). \doi{10.1117/12.411879}

\bibitem{harrington2004a}
Harrington, S.J., Naveda, J.F., Jones, R.P., Roetling, P., Thakkar, N.:
  Aesthetic measures for automated document layout. In: Munson, E.V.,
  Vion-Dury, J.Y. (eds.) DocEng '04: Proceedings of the 2004 ACM Symposium on
  Document Engineering. pp. 109--111. ACM (10 2004).
  \doi{10.1145/1030397.1030419}

\bibitem{levin2021a}
Levin, G., Brain, T.: Code as Creative Medium: A Handbook for Computational Art
  and Design. The MIT Press (2021)

\bibitem{lok2004a}
Lok, S., Feiner, S., Ngai, G.: Evaluation of visual balance for automated
  layout. In: Vanderdonckt, J., Nunes, N.J., Rich, C. (eds.) Proceedings of the
  9th international conference on Intelligent user interfaces. pp. 101--108.
  ACM (1 2004). \doi{10.1145/964442.964462}

\bibitem{lopes2023a}
Lopes, D., Correia, J., Machado, P.: Towards the automatic evaluation of visual
  balance for graphic design posters. In: Pease, A., Cunha, J.M., Ackerman, M.,
  Brown, D.G. (eds.) Proceedings of the 14th International Conference on
  Computational Creativity, ICCC 2023, Waterloo, June 19-23, 2023. pp.
  192--199. Association for Computational Creativity (6 2023)

\bibitem{lupton2014a}
Lupton, E.: Thinking with Type: A Critical Guide for Designers, Writers,
  Editors, \& Students. Princeton Architectural Press, 2nd edn. (2014)

\bibitem{machado2008a}
Machado, P., Romero, J., Manaris, B.: Experiments in Computational Aesthetics,
  pp. 381--415. Springer Berlin Heidelberg (2008).
  \doi{$10.1007/978-3-540-72877-1_18$}

\bibitem{meggs2016a}
Meggs, P.B., Purvis, A.W.: Meggs' History of Graphic Design. John Wiley \&
  Sons, Inc., 6th edn. (2016)

\bibitem{merelo2022a}
Merelo, J.J.: Agile (data) science: A (draft) manifesto. arXiv preprint
  arXiv:2104.12545  (7 2022). \doi{10.48550/arXiv.2104.12545}, (visited: 18
  January 2024)

\bibitem{mohammad2010a}
Mohammad, S., Turney, P.: Emotions evoked by common words and phrases: Using
  mechanical turk to create an emotion lexicon. In: Inkpen, D., Strapparava, C.
  (eds.) Proceedings of the NAACL HLT 2010 Workshop on Computational Approaches
  to Analysis and Generation of Emotion in Text. pp. 26--34. ACL (6 2010)

\bibitem{purvis2003a}
Purvis, L., Harrington, S., O'Sullivan, B., Freuder, E.C.: Creating
  personalized documents: an optimization approach. In: Proceedings of the 2003
  ACM Symposium on Document Engineering. pp. 68--77. ACM (8 2003).
  \doi{10.1145/958220.958234}

\bibitem{rebelo2021a}
Rebelo, S.M., Martins, T., Bicker, J., Machado, P.: Exploring automatic fitness
  evaluation for evolutionary typesetting. In: Sas, C., Maiden, N.A.M., Bailey,
  B.P., Latulipe, C., Do, E.Y.L. (eds.) C\&C '21: Creativity and Cognition
  Virtual Event Italy June 22 - 23, 2021. ACM (6 2021).
  \doi{10.1145/3450741.3465247}, (Article no. 12)

\bibitem{reynar1997a}
Reynar, J.C., Ratnaparkhi, A.: A maximum entropy approach to identifying
  sentence boundaries. In: Grishman, R. (ed.) Proceedings of the fifth
  conference on Applied natural language processing. pp. 16--19. ACL (3 1997).
  \doi{10.3115/974557.974561}

\bibitem{richardson2016a}
Richardson, A.: Data-driven Graphic Design: Creative Coding for Visual
  Communication. Bloomsbury Publishing Plc (2016)

\bibitem{runarsson2000a}
Runarsson, T.P., Yao, X.: Stochastic ranking for constrained evolutionary
  optimization. IEEE Transactions on Evolutionary Computation  \textbf{4},
  284--294 (9 2000). \doi{10.1109/4235.873238}

\bibitem{syswerda1989a}
Syswerda, G.: Uniform crossover in genetic algorithms. In: Schaffer, J.D. (ed.)
  Proceedings of the 3rd International Conference on Genetic Algorithms.
  pp.~2--9. Morgan Kaufmann Publishers Inc. (6 1989)

\bibitem{xie2021a}
Xie, Y., Huang, D., Wang, J., Lin, C.Y.: Canvasemb: Learning layout
  representation with large-scale pre-training for graphic design. In: MM'21:
  Proceedings of the 29th ACM International Conference on Multimedia. pp.
  4100--4108. ACM (10 2021). \doi{10.1145/3474085.3475541}

\end{thebibliography}

\end{document}